\documentclass[twocolumn, conference]{IEEEtran}

 \usepackage{graphicx}
\usepackage{amssymb}
\usepackage{booktabs}
\usepackage{textcomp}
 \usepackage{amsthm}
 \usepackage{amsmath}
 \usepackage{tabularx}
 \usepackage{placeins}
 \usepackage{amsfonts}
 \usepackage{placeins}
 \usepackage{lipsum}
 \usepackage{multicol}

\begin{document}
\title{ Wideband Parameters Analysis and Validation for Indoor radio Channel at 60/70/80GHz for Gigabit Wireless Communication employing Isotropic, Horn and Omni directional Antenna}

\author{\IEEEauthorblockN{}
\and
\IEEEauthorblockA{ E. Affum$ ^1 $ E.T. Tchao$ ^2 $  K. Diawuo$ ^3 $  K. Agyekum$ ^4 $ \\ Kwame Nkrumah Univ. of Science and Tech Kumasi, Ghana $ ^{1,2,3,4} $. \\   eaffume@gmail.com$ ^1 $,ettchao.coe@knust.edu.gh$ ^2 $, kdiawuo.soe@knust.edu.gh$ ^3 $, kwame.agyekum $ ^4 $}
}


\maketitle

\IEEEcompsoctitleabstractindextext{%
\begin{abstract}
Recently, applications of millimeter (mm) waves for high-speed broadband wireless local area network communication systems in indoor environment are increasingly gaining recognition as it provides gigabit-speed wireless communications with carrier-class performances over distances of a mile or more due to spectrum availability and wider bandwidth requirements. Collectively referred to as E-Band, the millimeter wave wireless technology present the potential to offer bandwidth delivery comparable to that of fiber optic, but without the financial and logistic challenges of deploying fiber. This paper investigates the wideband parameters using the ray tracing technique for indoor propagation systems with rms delay spread for Omni-directional and Horn Antennas for Bent Tunnel at 80GHz. The results obtained were 2.03 and 1.95 respectively, besides, the normalized received power with $0.55 \times 10^{8} $ excess delay at 70GHz for Isotropic Antenna was at 0.97.

\end{abstract}
\begin{IEEEkeywords}
Indoor; Wideband; Isotropic; rms Delay; Power delay Profile; Excess delay.
\end{IEEEkeywords}}

\IEEEdisplaynotcompsoctitleabstractindextext
\IEEEpeerreviewmaketitle

\section{Introduction}
With end users ranging from corporate data centers to teenagers with iPhones demanding higher bandwidth, the demand for newer technologies to deliver this bandwidth is higher than ever before. A plethora of technologies exist for the delivery of bandwidth, with fiber optic cable considered to be the ultimate bandwidth delivery medium. However, the fiber optics are not unmatched \cite{1} by any means, especially when all economic factors are considered. Millimeter wave wireless technology presents the potential to offer bandwidth delivery comparable to that of fiber optics, but without the financial and logistic challenges of deploying fiber. This paper is intended to analyze the wideband parameters of this new technology for different propagation indoor environment. Smulders studied wideband measurements of indoor radio channels operating in a 2 GHz frequency band \cite{2} centered around 58GHz using a frequency step sounding technique. The results were presented for cell coverage and root mean square (RMS) 
delay  spreads under both line-of-sight (LOS) and obstructed (OBS) situations. Again, various measurement campaigns and modeling activities were carried out \cite{3} to obtain both the narrowband as well as the wideband characteristics of the 60 GHz channel for indoor and outdoor environments. A simple ray-tracing was used to estimate the channel characteristics, for both narrow and wideband transmission systems in indoor as well as outdoor environments. Normalized received power, RMS delay profile and channel impulse response were simulated for indoor and outdoor radio channel. Ray tracing measurement of statistical parameters was comparatively studied and graphically represented by the group. Coherence bandwidth of wideband channel was estimated by Tlich and the group in \cite{4}. They further provided sounding measurements in the 30 kHz–100MHz band in several indoor environments. The coherence bandwidth and the RMS delay spread parameters were estimated from measurements of the complex transfer function 
and dispersion in 
the time domain, further, the variability of the coherence bandwidth and time-delay spread parameters with the channel class were presented based on the location of the receiver with respect to the transmitter and finally, related the RMS delay to the coherence bandwidth. 

In April 2007, A 60GHz indoor propagation channel model based on the ray-tracing method was proposed by Chong et al, and in that study, they validated the proposed model with measurements conducted in indoor environment \cite{5}. Moraitis and the group proposed the propagation models based on geometrical optics using ray-tracing theory for millimeter wave frequencies \cite{6}. Expressions for Path loss, Received power, Power delay profile (PDP) and RMS delay spread were presented by Moraitis and the group in \cite{7}. They performed propagation measurements at 60 GHz and determined the characteristics of indoor radio channels between fixed terminals that were illustrated. Path loss measurements were reported for line-of-sight (LOS), and non-line-of-sight (NLOS) cases, fading statistics in a physically stationary environment were extracted and effect due to the movement of the user on the temporal fading envelope were investigated. Path loss was predicted for models that provide excellent fitting with errors 
having dynamic 
range of fading in a quiescent environment.. In \cite{8}, a model for indoor radio propagation at 60GHz was used for predicting the performance of high speed wireless data networks, by using electromagnetic theory. Different models of the indoor environment were evaluated and corresponding received power, impulse response of the channel were predicted. Further, a statistical propagation model for the 60-GHz channel in a medium-sized room was presented by the Authors \cite{9}. Extensive work was done in \cite{10}. The  paper they proposed an area prediction system, which was capable of accurately predicting indoor service areas using the ray-tracing method, when a base station antenna was installed indoors. Ray tracing techniques was discussed in various models and units. Received power and delay spread were estimated and simulated. Moreover, there was an in depth discussion on 60-GHz radio in different aspects. Propagation and antenna effects were studied in line-of-sight and non-line-of sight environments 
and Bit error 
rate were simulated for different noise level. The selections of wideband channel sounding measurements were performed as part of the AWACS (ATM Wireless Access Communications System). The results were obtained for two different indoor operating environments (mainly in line-of-sight conditions) at a carrier frequency of 19.37 GHz.. This paper analyzed the wideband parameters using a proposed indoor propagation environment with dimensions of 4.4m $\times$ 2.5m representing a Straight Tunnel and a propagation environment using Bent Tunnel of angle of curvature of 45 with maximum height of 2m. Further, simulation results have been provided indicating normalized received power for different propagation environment. This paper is organized as follows: The propagation environments were first considered. Also the Narrowband and Wideband parameters were analyzed. This is followed by simulation results and discussion and finally conclusion.

\section{PROPAGATION ENVIRONMENT}
There are two general types of propagation modeling: site-specific and site general. Site-specific modeling requires detailed information on building layout, furniture, and transceiver locations \cite{11}. It is performed using ray-tracing methods. For large-scale static environments, this approach may be viable. For most environments, however, the knowledge of the building layout and materials is limited and the environment itself can change by simply moving furniture or doors. Thus, the site-specific technique is not commonly employed. Site-general models provide gross statistical predictions of path loss for link design and are useful tools for performing initial design and layout of indoor wireless systems. In this work site-specific indoor environment is considered. Different types of environment were considered, namely: Plain Corridor, Corridor with equally spaced wooden door, glass door and lift, Straight Tunnel and Bent Tunnel

\subsection{Plain Corridor}
\begin{figure*}
\centering
 \begin{tabular}{cc}
  \includegraphics{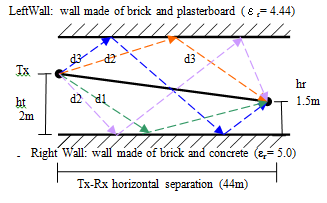} & \includegraphics{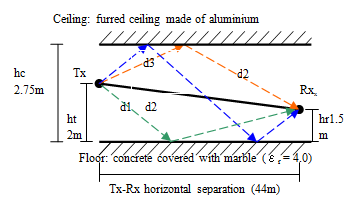} \\
  (a) & (b)  \\
 \end{tabular}
 \caption{Propagation Environment-Plain Corridor}
  \label{f1}
\end{figure*}

The propagation environment is a long plane corridor with dimensions $44 \times 2.20 \times 2.75 m^3$ as shown in Figure \ref{f1}. The left and right wall surfaces of the corridor are made of brick and plasterboard (relative permittivity $\varepsilon_r= 4.44 and \varepsilon_r= 5.0$). In order to simplify the simulation procedure, it was assumed that the surface is a uniform wall made of brick and plasterboard \cite{9}. The floor is made of concrete covered with marble ($\varepsilon_r= 4.0$) and furred ceiling is made of aluminum ($\varepsilon_r= 1.0$) as shown in Figure \ref{f1}

\subsection{Corridor with Wooden Door, Glass Door \& Lift}
\begin{figure*}
\centering
 \begin{tabular}{cc}
  \includegraphics[width=0.4\linewidth]{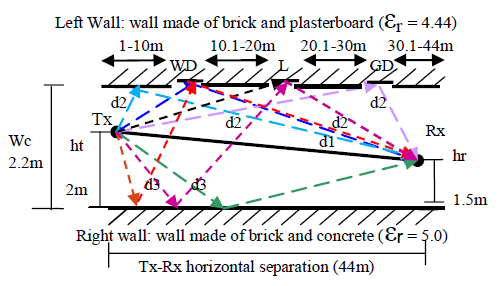} & \includegraphics[width=0.4\linewidth]{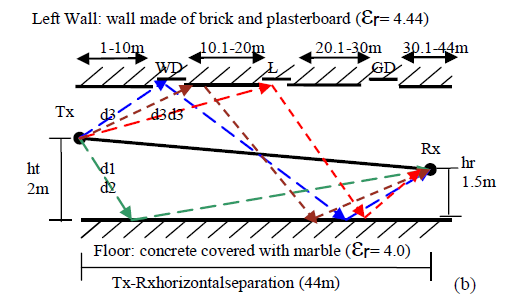} \\
  (a) & (b)  \\
 \end{tabular}
 \caption{Propagation Environment: Corridor with Wooden door, Glass and Lift  \cite{8}}
  \label{f2}
\end{figure*}

Figures \ref{f2} illustrate the corridor with wooden Door, Lift and Glass door. Between the distances of about 1-10m and 10.1-20m is a plane corridor, in between is a wooden door, similarly, in between 20-20.1m is a lift , and  30-30.1m glass door is present, from 30.1-40m is a plain corridor. Relative permittivity of wooden door is 3.3.

\subsection{Straight Tunnel}
\begin{figure*}
\centering
 \includegraphics[width=0.4\linewidth]{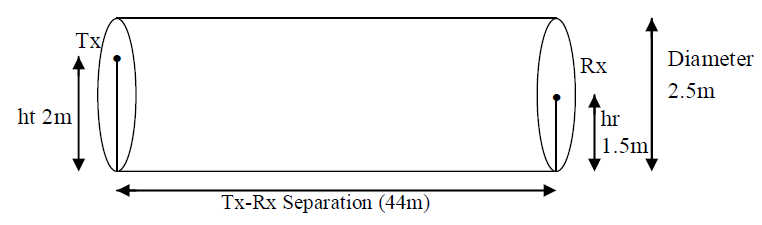}
 \caption{Straight Tunnel made of concrete}
 \label{f3}
\end{figure*}

The indoor propagation environment is a long tunnel with dimensions $44m \times 2.5 m$ as shown in the Figure \ref{f3}. The surfaces of the tunnel are made of concrete (relative permittivity $(\varepsilon_r)= 5.0$). The height of the transmitter is 2m and height of the receiver is 1.5m.

\subsection{Bent Tunnel}
 In figure 4, the propagation environment is a bent tunnel with dimensions $44 \times 2.5 m^2$.  The angle of curvature of the Bent tunnel is 45o. The surfaces of the tunnel are made of concrete (relative permittivity $(\varepsilon_r) = 5.0$). The height of the transmitter is 2m and that of the receiver is 1.5m.The electromagnetic rays from the transmitter gets reflected, refracted, diffracted, scattered and absorbed by the propagation environment before reaching the receiver. The received signal is the combination of all the signals. At 60/70/80 GHz, the diffraction phenomenon is almost negligible and the diffracted power does not contribute to the total received power. So the Diffraction was not taken into account. The non-uniformities of the surface materials in indoor environments are such that the produced scattering is not a substantial contribution to the received power thus up to second order reflected rays were taken into consideration, since further reflected rays i.e., third, fourth and so on, 
have insignificant 
contribution to the total received power. Atmospheric propagation losses were not taken into account since in indoor environments the attenuation is very small (0.00116 dB/m).
The beginning and the end of the corridor are open areas and were not taken into account in the simulations. The radio channel propagation modeling at millimeter wave frequencies can be realized based on ray-tracing theory. The ray-tracing method is among the available methods for the relatively accurate estimation of field strengths to deal with the type of complex layout that is often found in indoor environments. Ray-tracing allows fast computation of single and double reflection processes. In the 60/70/80 GHz region the diffraction phenomenon can be neglected and the sum of the direct ray and the reflected rays are enough to describe the behavior of the propagation channel with great accuracy.

\section{NARROWBAND PARAMETERS}

\subsection{Received Power}
The total received power (RR) of the multi-rays are calculated by \cite{6} the summation of `x’ single reflected and `w’ double reflected rays given by

\begin{equation}
\begin{split}
R_{R} & = T_{R} \left( \frac{\lambda}{4 \pi} \right)^{2} a_{t}a_{r} \left| \frac{e^{-jkd_{1}}}{d_{1}} + \sum_{i=1}^{x} R(\theta_{0}) \frac{e^{-jkd_{2}}}{d_{2}} \right. \\ 
&  \left.+  \sum_{j=1}^{w} R(\theta_{1})R(\theta_{2}) \frac{e^{-jkd_{3}}}{d_{3}} \right|^{2}   \label{e1}  
 \end{split}
\end{equation}

where $\lambda$ is the wave length; $k$ is the wave number, $d_1$ is the distance of the direct path; $d_2$ is the distance of the single reflected path; $d_3$ is the distance of the double reflected path; $a_t, a_r$ are the antenna functions;     $R(\theta_{0}) $ is the reflection coefficient of the single reflected ray on the reflecting surface; $R(\theta_{1})R(\theta_{2})$ are the reflection coefficient of the double reflected rays on respective reflecting surfaces; and $T_R$ is the transmitted power. For isotropic antennas ($a_t = a_r = 1$) the total received power \cite{12} ($R_R$) is

\begin{equation}
\begin{split}
R_{R} & = T_{R} \left( \frac{\lambda}{4 \pi} \right)^{2}  \left| \frac{e^{-jkd_{1}}}{d_{1}} + \sum_{i=1}^{x} R(\theta_{0}) \frac{e^{-jkd_{2}}}{d_{2}} \right. \\ 
&  \left.+  \sum_{j=1}^{w} R(\theta_{1})R(\theta_{2}) \frac{e^{-jkd_{3}}}{d_{3}} \right|^{2}   \label{e1}  
 \end{split}
\end{equation} 

To examine the signal propagation in the indoor environment, we assumed three different transmission systems with different antenna characteristics and transmitted power. This was done so as to examine how the antenna radiation patterns affect the signal propagation in the indoor environment. The systems are: System 1: Isotropic antennas on both transmitter and receiver and 20 dBm output power. System 2: Transmitter power = 20 dBm, Transmitter Gain = Receiver Gain = 8.5 dBi for omnidirectional antenna System 3: Transmitter power = 10 dBm, Transmitter Gain = Receiver Gain = 20.8dBi also for horn antenna. Finally, the simulation was conducted with MATLAB script, using multi rays. The initial transmitter position is at the beginning of the propagation environment and the receiver is moving at almost constant speed. The total number of samples for the entire propagation environment (44 m) was 1024.

\section{WIDEBAND PARAMETERS}

\subsection{Power Delay Profile}
The wideband multipath channel is often modeled as a time varying linear filter with complex impulse response \cite{12}

\begin{equation}
 (t, \tau) =  \sum_{i=1}^{N} a_{t} (t, \tau) exp (j \psi_{i}(t, \tau) \delta (\tau - \tau_{i}))
\end{equation}

For the indoor propagation environment where the time varying factors of the impulse response typically are human movement, it is appropriate to treat the channel as quasi-stationary. Assuming that the phase variations in the CIR have a uniform distribution we may consider only the amplitude and the delay components. The most significant parameter derived from the procedure is the received power as a function of the time delay known also as the power delay profile (PDP). The power delay profile can be

\begin{equation}
 P(\tau) = \sum\limits_{i=1}^{N} P_{r}(d) \delta(\tau - \tau_{i})
 \label{e4}
\end{equation}

where, $P_r(d)$ is the received signal of the $i^th$ ray and N is the total number of the rays used in the simulation procedure. The average received power in every bin is normalized to the maximum received power. The examination of the signal propagation in the indoor environment, using three different transmission systems with different antenna characteristics and transmitted power were considered. This was done so as to examine how the antenna

\begin{figure*}
\centering
 \includegraphics{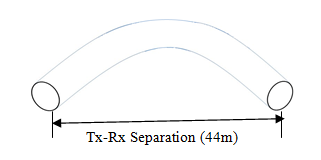}
 \caption{Bent Tunnel made of concrete}
 \label{f4}
\end{figure*}

\begin{figure*}
\centering
\begin{minipage}{.5\textwidth}
  \centering
\includegraphics[width= 0.9\textwidth, height= 0.7\textwidth]{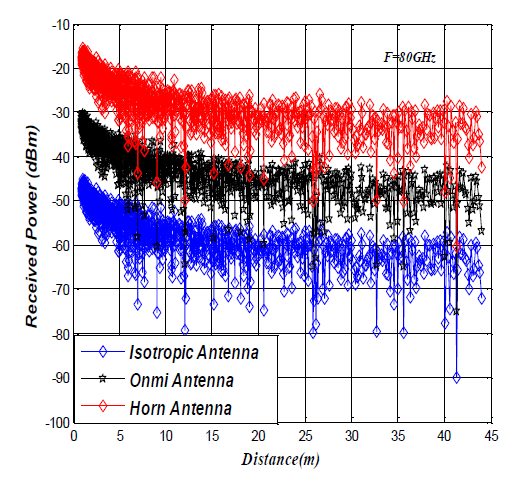}
\caption{ Received power for different antenna systems at 80 GHz with Corridor with
Wooden Doors, Glass Door and Lift}
\label{f5}
\end{minipage}%
\begin{minipage}{.5\textwidth}
  \centering
\includegraphics[width= 0.9\textwidth, height= 0.7\textwidth]{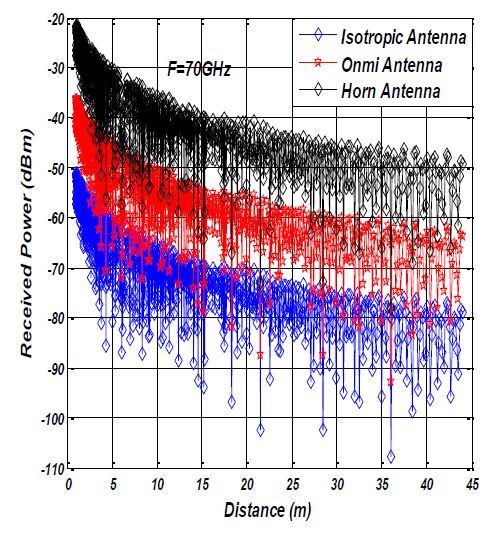}
\caption{Received power for different antenna system configurations at 70 GHz for Straight tunnel}
\label{f6}
\end{minipage}
\end{figure*}

\begin{figure*}
\centering
\begin{minipage}{.5\textwidth}
  \centering
\includegraphics[width= 0.9\textwidth, height= 0.7\textwidth]{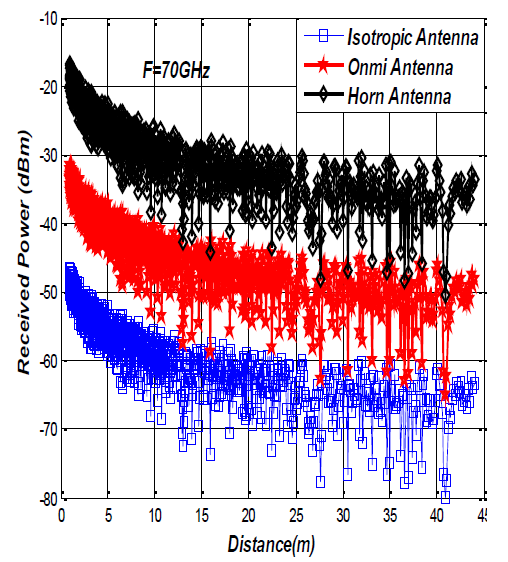}
\caption{ Received power for different antenna system configurations at 70 GHz for Bent Tunnel} 
\label{f7}
\end{minipage}%
\begin{minipage}{.5\textwidth}
  \centering
\includegraphics[width= 0.9\textwidth, height= 0.7\textwidth]{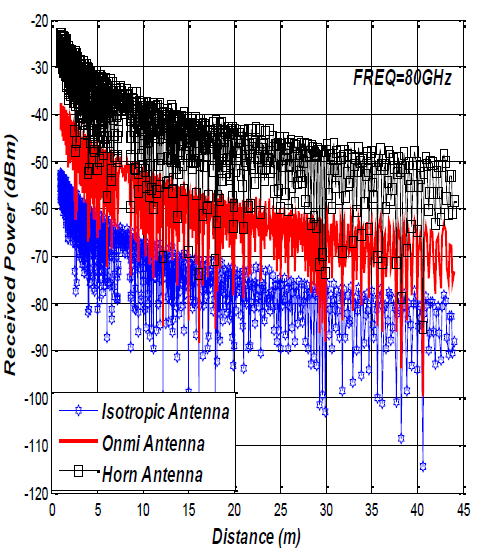}
\caption{Received power for different antenna system configurations at 80 GHz for Bent tunnel}
\label{f8}
\end{minipage}

\end{figure*}

\begin{figure*}
\begin{minipage}{.5\textwidth}
  \centering
\includegraphics[width= 0.9\textwidth, height= 0.7\textwidth]{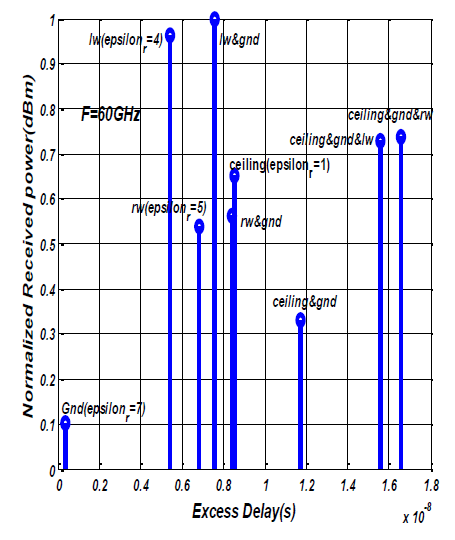}
\caption{Delay Spread of Straight Tunnel } 
\label{f9}
\end{minipage}%
\begin{minipage}{.5\textwidth}
  \centering
\includegraphics[width= 0.9\textwidth, height= 0.7\textwidth]{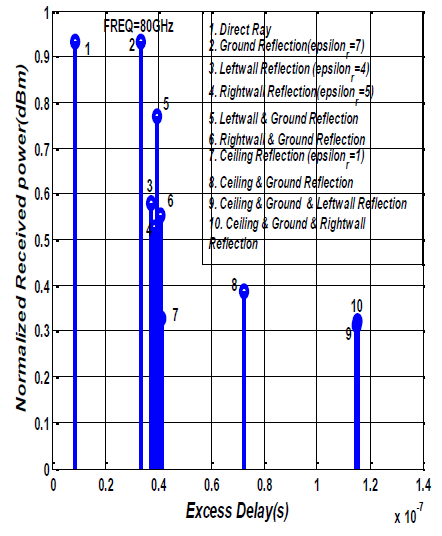}
\caption{Delay Spread of Straight Tunnel}
\label{f10}
\end{minipage}
\end{figure*}


\begin{table}[!ht]
 \centering
 \caption{rms Delay Spread of Straight Tunnel}
 \label{t1}
 \begin{tabular}{cccc}
 \hline
Antennas & 60 GHz (ns) & 70 GHz (ns) & 80 GHz (ns)\\
\hline
Isotropic & 1.92 & 1.93 & 1.94\\
Omni & 1.84 & 1.85 & 1.86 \\
Horn & 1.77 & 1.78 &  1.79 \\
\hline
 \end{tabular}

\end{table}

\begin{table}[!ht]
 \centering
 \caption{rms Delay Spread of Bent  Tunnel}
 \label{t2}
 \begin{tabular}{cccc}
 \hline
Antennas & 60 GHz (ns) & 70 GHz (ns) & 80 GHz (ns)\\
\hline
Isotropic & 2.24 & 2.25 & 2.26\\
Omni & 2.14 & 2.15 & 2.16\\
Horn & 2.05 & 2.06 & 2.07 \\
\hline
 \end{tabular}

\end{table}

\section{RESULTS AND DISCUSSIONS}
Radiation patterns affect the signal propagation in the indoor environment. For Isotropic antennas both transmitter and receiver output power was 20 dBm with unity gain, for Omni-directional antenna the transmitter power was also 20 dBm with transmitter and receiver gain as 8.5 dBi, besides, the transmitter power for horn antenna was 10 dBm with transmitter and receiver gain as 20.8 dBi. The height of the transmitter and receiver antennas in all propagation environments were 2m and 1.5m respectively. The distance between the transmitter and receiver was 44m and the total number of samples for the entire propagation environment (44 m) to be 1024.Figure \ref{f5}, \ref{f6}, \ref{f7} and \ref{f8} illustrate the total received power for different antenna system configurations under different propagation scenarios. It was observed that the received power of Horn antenna in Plain corridor at distance 10m was -30dBm, and Omni antenna were -45dBm and -60dBm for Isotropic antenna. Also, the received power for Plain 
Corridor at 70GHz at a distance 
40m, was-30dBm, the received power for Horn antenna is -35dBmand for Omni antenna the received power was also observed as -50dBm and -60dBm approximately for Isotropic antenna. Again for Plain Corridor at 80GHz. for a distance 20m, the received power for Horn antenna -35dBm and the received power for Omni antenna is -50dBm and -65dBm for Isotropic antenna approximately. The received power for Corridor with wooden door, Lift and Glass door for 60GHz frequency at a distance of 10m, of Horn antenna was -30dBm, for Omni and Isotropic antennas were -40dBm and -55dBm approximately. At Frequency 70GHz, the received power of Horn antenna at a distance of 20m was -30dBm, for Omni and Isotropic antenna the received powers -45dBm and 60dBm respectively. At 80GHz frequency, the received power of Horn antenna at a distance 30m, was -35dBm, for Omni antenna the received power is -50dBm. The received power in a Bent Tunnel at 60GHz was analyzed, at distance 10m, the power for Horn antenna was -35dBm, Omni antenna as 
50dBm and -65dBm for Isotropic antenna. Also the for Bent Tunnel at 70GHz as shown in figure \ref{f7},  for a distance of 20m, the received power for Horn antenna is -40dBm, and for Omni antenna the received power is -55dBm and -70dBm approximately for Isotropic antenna. The Power Delay Profile of Horn antenna at 60/70/80 GHz in Plain Corridor was also analyzed. The normalized received power of ceiling and ground reflected ray was 0.438 dBm with excess delay of $0.0724 \times 10^{-6}s$ and the normalized received power of ceiling, ground and left wall reflected ray was 0.364dBm with excess delay of $0.1152 \times 10^{-6}s$. 
Figure \ref{f10}, illustrates the Power Delay Profile of Isotropic antenna at 70GHz. The Normalized received power of right wall and ground reflected ray was 0.527dBm with excess delay of $0.0386 \times 10^{-7}s$ and the normalized received power of ceiling reflected ray was 0.386dBm with excess delay of $0.0723 \times 10^{-7}s$ approximately. The normalized received power of direct ray at 80GHz is 0.931dBm with excess delay of $0.0333 \times 10^{-7}s$ and 0.818dBm was the normalized received power of ceiling and ground reflected ray with excess delay of$ 0.041 \times 10^{-7}s$.

\subsection{Conclusion}
This paper has presented the characteristics of the propagation channel at 60/70/80 GHz, using four different indoor environments namely: Plain Corridor, Corridor with wooden door, lift and glass door, Straight Tunnel and Bent Tunnel employing Horn Antenna, Isotropic Antenna and Omni Antenna, utilizing MATLAB. From the results analyzed so far the Horn antenna performance outweighs the Omni antenna and Isotropic antenna in all indoor environments at 60/70/80 GHz. The rms delay spread of propagation environments obtained also indicated that as the frequency increases, the rms delay spread of the propagation environment increased gradually

\bibliographystyle{unsrt}
\bibliography{tutuu}

\end{document}